\documentclass[aps,prl,amsmath,preprint,groupedaddress]{revtex4-1}
\usepackage{graphicx}
\usepackage{hyperref}
\usepackage{multirow}
\usepackage{booktabs}

\begin{document}


\title{Searching for a possible dipole anisotropy on acceleration scale with 147 rotationally supported galaxies}



\author{Zhe Chang$^{1,2}$}
\author{Hai-Nan Lin$^3$}
\author{Zhi-Chao Zhao$^{1,2}$}
\author{Yong Zhou$^{1,2}$}
\email{zhouyong@ihep.ac.cn}
\affiliation{$^1$Institute of High Energy Physics, Chinese Academy of Sciences, Beijing 100049, China \\
$^2$School of Physical Sciences, University of Chinese Academy of Sciences, Beijing 100049, China\\
$^3$Department of Physics, Chongqing University, Chongqing 401331, China}

\date{\today}

\begin{abstract}

We report a possible dipole anisotropy on acceleration scale $g_{\dag}$ with 147 rotationally supported galaxies in local Universe. It is found that a monopole and dipole correction for the radial acceleration relation can better describe the SPARC data set. The monopole term is negligible but the dipole magnitude is significant.  It is also found that the dipole correction is mostly induced by the anisotropy on the acceleration scale. The magnitude of $\hat{g}_{\dag}$-dipole reaches up to $0.25\pm0.04$, and its direction is aligned to $(l,b) = (171.30^{\circ}\pm7.18^{\circ}, -15.41^{\circ}\pm4.87^{\circ})$, which is very close to the maximum anisotropy direction from the hemisphere comparison method. Furthermore, robust check shows that the dipole anisotropy couldn't be reproduced by isotropic mock data set. However, it is still premature to claim that the Universe is anisotropic due to the small data samples and uncertainty in the current observations.

\end{abstract}


\maketitle


\section{1. Introduction}
\label{Introduction}

One of the foundations of the standard cosmological paradigm ($\Lambda$CDM) is the so-called cosmological principle, which states that the Universe is homogeneous and isotropic on large scales \citep{Weinberg:2008zzc}. This principle is in accordance with most cosmological observations, especially with the approximate isotropy of the cosmic microwave background (CMB) radiation from Wilkinson Microwave Anisotropy Probe (WMAP) \citep{Bennett:2012zja,Hinshaw:2012aka} and Planck satellites \citep{Ade:2013zuv,Ade:2015xua}. However, there still exists some cosmological observations that challenge the cosmological principle. These include the large-scale alignments of quasar polarization vectors \citep{Hutsemekers:2005iz}, the unexpected large-scale bulk flow \citep{Kashlinsky:2009dw,Feldman:2009es}, the spatial variation of the fine structure constant \citep{Webb:2010hc,King:2012id}, and the dipole of supernova distance modulus \citep{Mariano:2012wx,Chang:2014wpa,Chang:2014nca,Lin:2016jqp}. All of these facts hint that the Universe may be anisotropic to some extent.

In galactic scale, the mass discrepancy problem \citep{Rubin:1978kmz,Bosma:1981zz} has been found for many years. The observed gravitational potential cannot be explained by the luminous matter (stellar and gas). Hence it seems that there needs a significant amount of non-luminous matter, i.e. the dark matter in galaxy system. But up to now, no direct evidence of the existence of dark matter has been found \citep{Tan:2016zwf,Aprile:2016swn}. A successful alternative to the dark matter hypothesis is the modified Newtonian dynamics (MOND) \citep{Milgrom:1983ca}, which attributes the mass discrepancies in galactic systems to a departure from the standard dynamics at low accelerations.

In principle, the MOND theory assumes a universal constant acceleration scale for all galaxies \citep{Milgrom:1983ca,Milgrom:2002tu,Milgrom:2012xw}. But in practice, the acceleration scale is considered as a free parameter to fit the galaxy rotation curve, and different galaxies may have different acceleration scales \citep{Begeman:1991iy,Swaters:2010qe,Chang:2013twa}. \citet{Milgrom:1998aj} also suggested that the acceleration scale may be a fingerprint of cosmology on local dynamics and related to the Hubble constant. Therefore the cosmological anisotropy on large scales may imprint on the acceleration scale in local Universe. These ideas inspire us to investigate the possibility of spatial anisotropy on the acceleration scale. In our previous work \citep{Zhou:2017lwy}, by making use of the hemisphere comparison method to search for such an anisotropy from the SPARC data set, we found that the maximum anisotropy level is significant and reached up to $0.37\pm0.04$ in the direction $(l,b) = ({175.5^\circ}^{+6^\circ}_{-10^\circ}, {-6.5^\circ}^{+9^\circ}_{-3^\circ})$. In this paper, we search a monopole and dipole correction for the radial acceleration relation, and try to find the possible anisotropy from the SPARC data set.

The rest of this paper is organized as follows. In Section 2, we make a brief introduction to the SPARC data set and the radial acceleration relation. In Section 3, we show a monopole and dipole correction for the radial acceleration relation by making use of the Markov chain Monte Carlo (MCMC) method to explore the entire parameter space. The information criterion (IC) is used for model comparison. In Section 4, the MCMC results for whole parameter spaces are analyzed, we compare the dipole anisotropy and  the goodness of fit for different dipole models. We also compare the possible dipole anisotropy with the hemisphere anisotropy in our previous work \citep{Zhou:2017lwy}. In Section 5, we make a robust check to examine whether the dipole anisotropy could be reproduced by a statistically isotropic data set. Finally, conslusions and discussions are given in Section 6.

\section{2. The data and radial acceleration relation}
\label{Data}

\begin{figure*}
\begin{center}
\includegraphics[width=\textwidth]{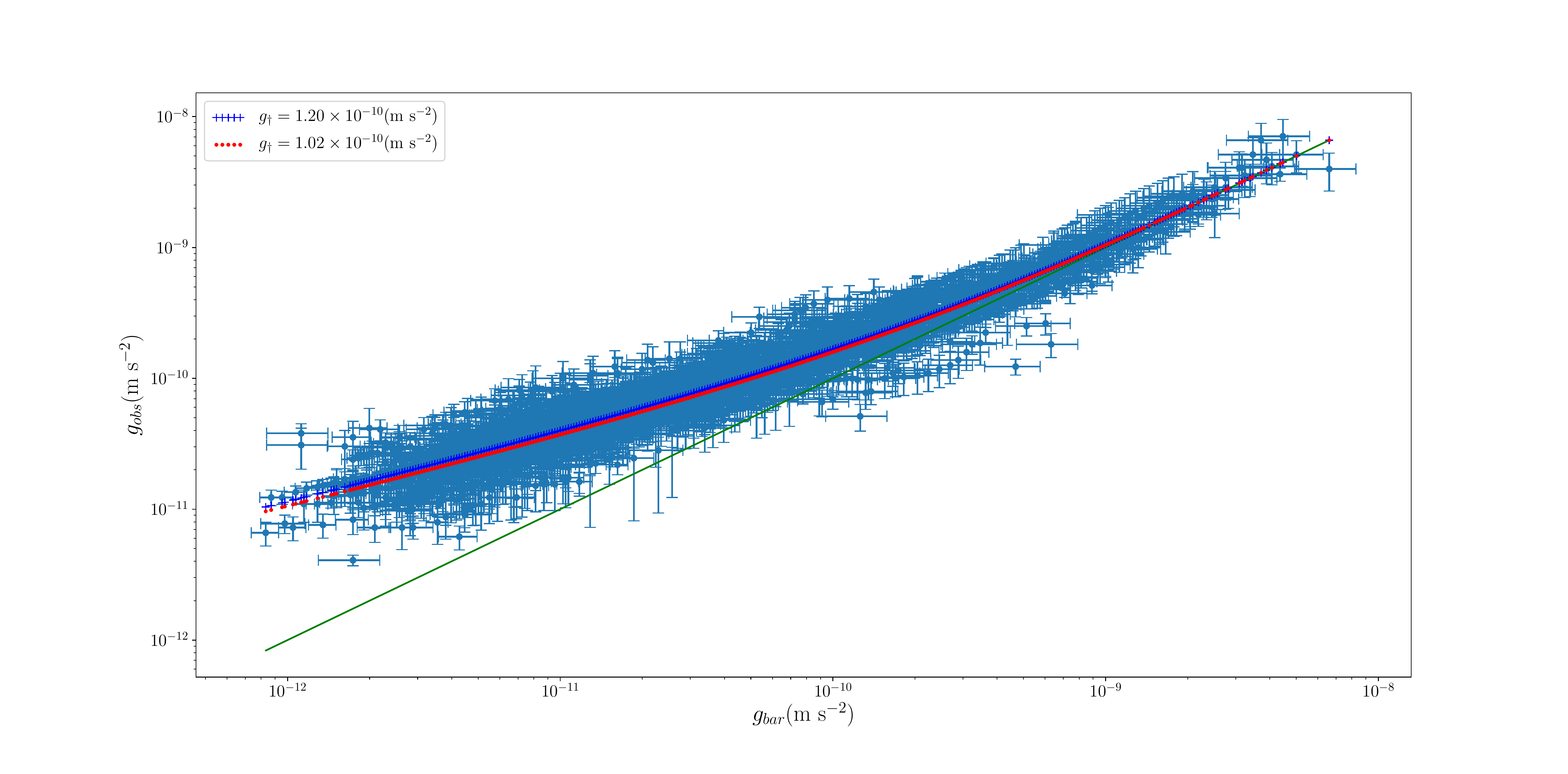}
\caption{The radial acceleration relation between the centripetal acceleration $g_{obs}$ and the baryonic acceleration $g_{bar}$ for all 2693 data points in 147 galaxies. Two dotted lines correspond to the fitting curve ($g_{\dag}=\rm 1.20\times 10^{-10}~m~s^{-2}$) in \citet{McGaugh:2016leg} and ours fitting curve ($g_{\dag}=\rm 1.02\times 10^{-10}~m~s^{-2}$), respectively. The solid line is the line of unity.}\label{fig:SPARC}
\end{center}
\end{figure*}

We employ the new Spitzer Photometry and Accurate Rotation Curves (SPARC) data set \footnote{\url{http://astroweb.cwru.edu/SPARC/}} \citep{Lelli:2016zqa}. The SPARC is a sample of 175 disk galaxies with new surface photometry at $3.6~\mu$m and high-quality rotation curves from previous HI/H$\alpha$ studies. For investigating the radial acceleration relation, \citet{McGaugh:2016leg} have adopted a few modest quality criteria to exclude some unreliable data. Finally, a sample of 2693 data points in 147 galaxies have been left. Here we use the same sample to search for the possible spatial dipole anisotropy. The SPARC data set does not include the galactic coordinate. We complete it for each galaxy from previous literatures \citep{Begum:2004sj,deBlok:1996jib} and by retrieving the NED dataset \footnote{\url{http://ned.ipac.caltech.edu/}}.

\citet{McGaugh:2016leg} obtained a fitting function that described well the radial acceleration relation for all 2693 data points. The fitting function is of the form
\begin{eqnarray}\label{eq:nuy}
g_{obs}=\frac{g_{bar}}{1-e^{-\sqrt{g_{bar}/g_{\dag}}}},
\end{eqnarray}
where $g_{bar}$ is the baryonic (gravitational) acceleration predicted by the distribution of baryonic mass and  $g_{obs}$ is the observed dynamic centripetal acceleration traced by rotation curves. The founction has a unique fitting parameter $g_{\dag}$, which corresponds to the MOND acceleration scale. They found $g_{\dag}=\rm [1.20\pm0.02~(random)\pm0.24~(systematic)]\times10^{-10}~m~s^{-2}$. This value is consistent with that predicted by the MOND theory \citep{Milgrom:1983ca,Milgrom:2016uye}. The MOND theory also predicts two limiting cases for the radial acceleration relation. In the deep-MOND limit, i.e.  $g_{bar}\ll g_{\dag}$, the fitting function \eqref{eq:nuy} becomes $g_{obs} \approx \sqrt{g_{bar} g_{\dag}}$, where the mass discrepancy appears. In the Newton limit, i.e. $g_{bar}\gg g_{\dag}$, the fitting function \eqref{eq:nuy} becomes $g_{obs} \approx g_{bar}$ and the Newtonian dynamics is recovered. The radial acceleration relation and the SPARC data points are illustrated in Fig. \ref{fig:SPARC}.

\section{3. Methodology}
\label{Methodology}
\subsection{3.1. The fitting method}
\label{Method}

As the same as \citet{McGaugh:2016leg}, we make use of the orthogonal-distance-regression (ODR) algorithm \citep{boggs1987stable} to fit the radial acceleration relation. The advantage of this method is that it could consider errors on both variables. The chi-square is defined as
\begin{eqnarray}\label{eq:chi}
\chi^2=\sum^n_{i=1}\frac{[g_{th}(g_{bar,i}+\delta_i,g_{\dag})-g_{obs,i}]^2}{\sigma^2_{obs,i}}+\frac{\delta_i^2}{\sigma^2_{bar,i}},
\end{eqnarray}
where $\sigma_{obs}$ and $\sigma_{bar}$  are the uncertainty of $g_{obs} $ and $g_{bar}$, respectively. The total number of data points is $n=2693$.  $\delta_i$ is an interim parameter which is used for finding out the weighted orthogonal (shortest) distance from the curve $g_{th}(g_{bar},g_{\dag})$ to the $i$th data point. The curve is same as the right-hand side of the function \eqref{eq:nuy}, i.e.
\begin{eqnarray}\label{eq:gth}
g_{th}(g_{bar},g_{\dag})=\frac{g_{bar}}{1-e^{-\sqrt{g_{bar}/g_{\dag}}}},
\end{eqnarray}
where $g_{th}$ represents the theoretical centripetal acceleration. Therefore, the chi-square \eqref{eq:chi} is the sum of the squares of the weighted orthogonal distances from the curve to the $n$ data points. Eventually, we minimize the chi-square to find out the best fitting value of $g_{\dag}$. We repeat the fitting process in \citet{McGaugh:2016leg} and reproduce the same result, if the logarithmic distance (base 10) of the first term in the chi-square \eqref{eq:chi} was taken. For the original form of chi-square \eqref{eq:chi}, we find the best fitting value for the universal acceleration scale is $g_{\dag}=\rm (1.02 \pm 0.02)\times10^{-10}~m~s^{-2}$, which corresponds to the unnormalized chi-square $\chi^2=4020$. This fitting curve is also plotted in Fig. \ref{fig:SPARC}. It is worth noting that the difference in the best fitting value comes from the form of the first term in the chi-squre \eqref{eq:chi}, which could only have slight impact on the possible dipole anisotropy. What we concern here is the relative variation of the acceleration scale in different direction, which is similar to the hemisphere anisotropy.

\subsection{3.2. The monopole and dipole correction}
\label{Model}

The MOND theory assumes a universal constant acceleration scale for all galaxies \citep{Milgrom:1983ca,Milgrom:2002tu,Milgrom:2012xw}. However, the acceleration scale has been found to be variational from galaxies to galaxies \citep{Begeman:1991iy,Swaters:2010qe,Chang:2013twa}. In our previous work \citep{Zhou:2017lwy}, we have found that there exists a hemisphere anisotropy on acceleration scale in the SPARC data set. In this paper, we show a monopole and dipole correction for the radial acceleration relation to search for a possible spatial dipole anisotropy in local Universe. The monopole and dipole correction is a commonly used method to search for possible dipole anisotropy, for instance, the spatial variation of the fine structure constant \citep{Webb:2010hc,King:2012id}, and the dipole of supernova distance modulus \citep{Mariano:2012wx,Chang:2014wpa,Chang:2014nca,Lin:2016jqp}. Here, we first assume the dipole anisotropy coming from the radial acceleration.
 The theoretical centripetal acceleration with a monopole and dipole correction is of the form
\begin{eqnarray}\label{eq:data}
\hat{g}_{th}(g_{bar},g_{\dag}) = g_{th}(g_{bar},g_{\dag})[1+A+B\hat{\textsl{\textbf{m}}}\cdot\hat{\textsl{\textbf{p}}}],
\end{eqnarray}
where the acceleration scale have been fixed at the best fitting value $g_{\dag}=\rm 1.02\times 10^{-10}~m~s^{-2}$.
$A$ and $B$ are the monopole term and dipole magnitude, respectively. $\hat{\textsl{\textbf{m}}}$ and $\hat{\textsl{\textbf{p}}}$ are the unit vectors pointing towards the dipole direction and galaxy position, respectively. In the galactic coordinates, the dipole direction can be presented as $\hat{\textsl{\textbf{m}}}=\cos(b)\cos(l)\hat{\textsl{\textbf{i}}}+\cos(b)\sin(l)\hat{\textsl{\textbf{j}}}+\sin(b)\hat{\textsl{\textbf{k}}}$, where $l$ and $b$ are galactic longitude and latitude, respectively. Similarly, the position of the $i$th galaxy can be presented as $\hat{\textsl{\textbf{p}}}_i=\cos(b_i)\cos(l_i)\hat{\textsl{\textbf{i}}}+\cos(b_i)\sin(l_i)\hat{\textsl{\textbf{j}}}+\sin(b_i)\hat{\textsl{\textbf{k}}}$. Then we use the corrected theoretical centripetal acceleration $\hat{g}_{th}(g_{bar},g_{\dag})$ for the chi-square \eqref{eq:chi}, and employ the MCMC method to explore the entire parameter space $\{A,B,l,b\}$. We don't have any information about the dipole anisotropy, thus a flat prior for the parameter space is needed, which will be discussed in Section 4. Actually, the MCMC method used here is same as the maximum likelihood method, the best fitting value corresponds to the minimal chi-square ($\chi^2_{min}=-2\ln \mathcal{L}_{max}$).

Second, we assume that the dipole anisotropy on radial acceleration is induced by spatial variation of the acceleration scale which is a unique parameter in radial acceleration relation. The acceleration scale with a monopole and dipole correction is of the form
\begin{eqnarray}\label{eq:parameter}
\hat{g}_{\dag} = g_{\dag}(1+C+D\hat{\textsl{\textbf{n}}}\cdot\hat{\textsl{\textbf{p}}}),
\end{eqnarray}
where the fiducial acceleration scale has also been fixed at the best fitting value $g_{\dag}=\rm 1.02\times 10^{-10}~m~s^{-2}$ and other parameters have analogous meanings with that in equation \eqref{eq:data}. Then we substitute the corrected acceleration scale \eqref{eq:parameter} into the theoretical centripetal acceleration \eqref{eq:gth}. By making use of $g_{th}(g_{bar},\hat{g}_{\dag})$ for the chi-square \eqref{eq:chi}, We employ the MCMC method to explore the entire parameter space $\{C,D,l,b\}$.  Finally we minimize the chi-square to obtain the best fitting value.

The monopole term in both corrections are retained for complete description, but usually the monopole term is negligible. As a contrast, we repeat the above process with only the dipole term for both corrections. Totally we have four corrections for the radial acceleration relation.

\subsection{3.3. Model comparison}
\label{IC}

To assess the goodness of fit and take account of the number of free parameters in each model, we employ the information criteria (IC) to compare the corrected model i.e. $\hat{g}_{th}(g_{bar},g_{\dag})$ or $g_{th}(g_{bar},\hat{g}_{\dag})$ with the reference model $g_{th}(g_{bar},g_{\dag})$. Here the corrected model could degenerate to the reference model when the monopole term and the dipole magnitude both equal to zero. Two most widely used information criteria are the Akaike information criterion (AIC) \citep{1974ITAC...19..716A} and the Bayesian information criterion (BIC) \citep{1978AnSta...6..461S}. They are defined as
\begin{eqnarray}\label{eq:IC}
\textrm{AIC}=\chi^2_{min}+2k,\\
\textrm{BIC}=\chi^2_{min}+k\ln n,
\end{eqnarray}
where $\chi^2_{min}$ is the minimal chi-square calculated by equation \eqref{eq:chi}, $k$ is the number of free parameters, and the total number of the data points is $n=2693$. Differently from AIC, due to $\ln 2693>2$, the BIC heavily penalizes models with the excess of free parameters. It is noteworthy that only the relative value of IC between different model is important in the model comparison. By convention, the model with $\Delta\textrm{IC}>5$ is regarded as `strong' and $\Delta\textrm{IC}>10$ as `decisive' evidence against the weaker model with higher IC value \citep{Liddle:2007fy,Arevalo:2016epc,Lin:2017yfl}.

\section{4. Result}
\label{Result}

We implement the MCMC method by using the affine-invariant Markov chain Monte Carlo ensemble sampler in $emcee$ \citep{2013PASP..125..306F}, which is adopted widely in the astrophysics and cosmology. One hundred random walkers are used to explore the entire parameter space. We run 500 steps in the burn-in phase and another 2000 steps in the production phase, which is enough for our purpose.  The dipole direction and its opposite direction is same for the correction when the dipole magnitude changes its sign. For obtaining unanimous result, we confine the dipole magnitude to be positive, and constrain the dipole direction at one cycle range, i.e. $l\in[0^\circ,360^\circ],~b\in[-90^\circ,90^\circ]$. The MCMC method needs a prior distribution for each parameter but we don't have any information about the dipole anisotropy. In this paper, we adopt a flat prior distribution for each parameter as follow: $A(C)\sim[-1,1],~B(D)\sim[0,1],~l\sim[0^\circ,360^\circ],~b\sim[-90^\circ,90^\circ]$.

\begin{figure}
\begin{center}
\includegraphics[width=8.6cm]{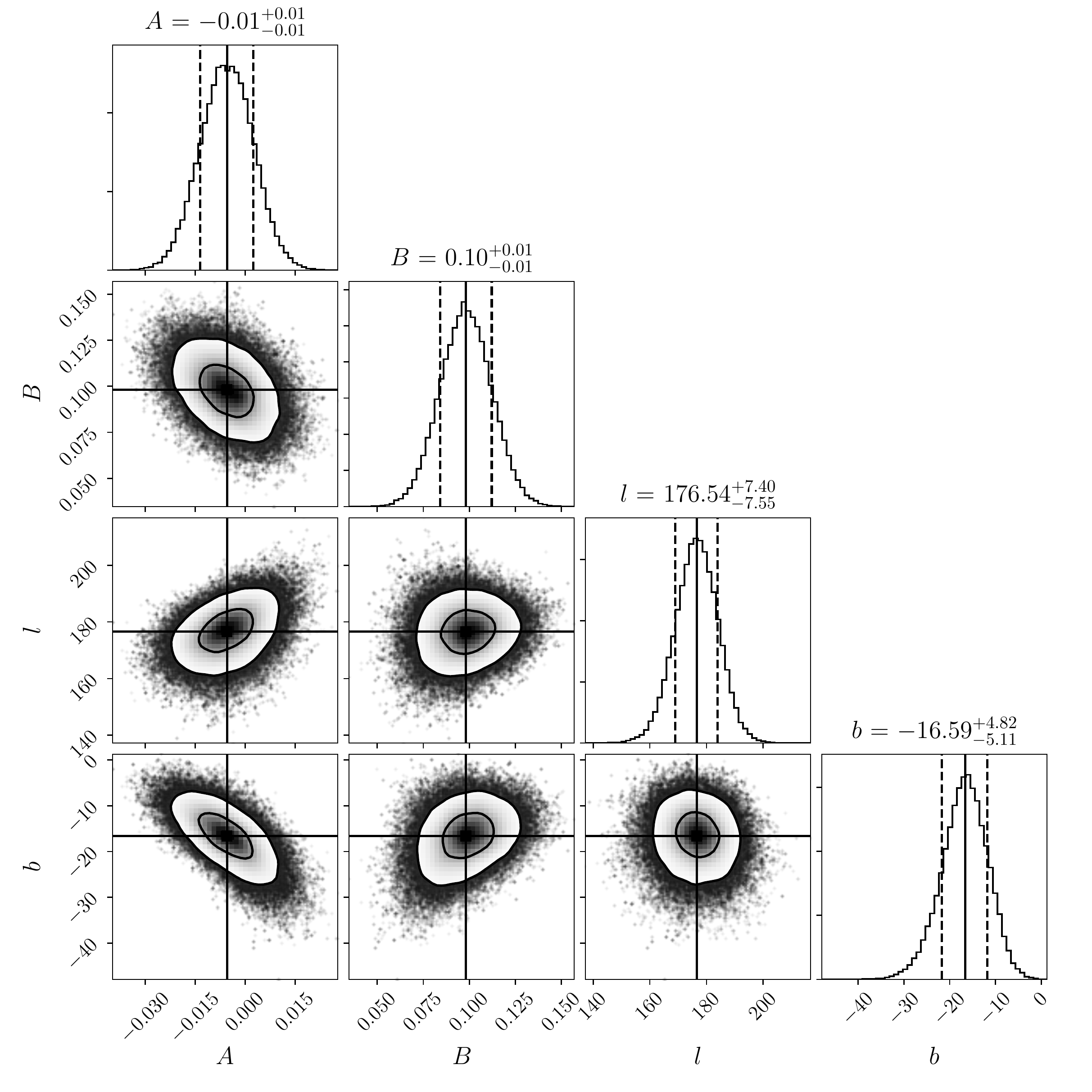}
\caption{The 1-dimensional marginalized histograms and 2-dimensional marginalized contours for the parameters space $\{A,B,l,b\}$. The horizontal and vertical solid lines mark the median values. The vertical dashed lines mark the $1\sigma$ credible intervals. These values are labeled at the top of each histogram. The 2-dimensional marginalized contours mark $1\sigma,~2\sigma$ credible regions from grey to light.}\label{fig:data}
\end{center}
\end{figure}

\begin{figure}
\begin{center}
\includegraphics[width=8.6cm]{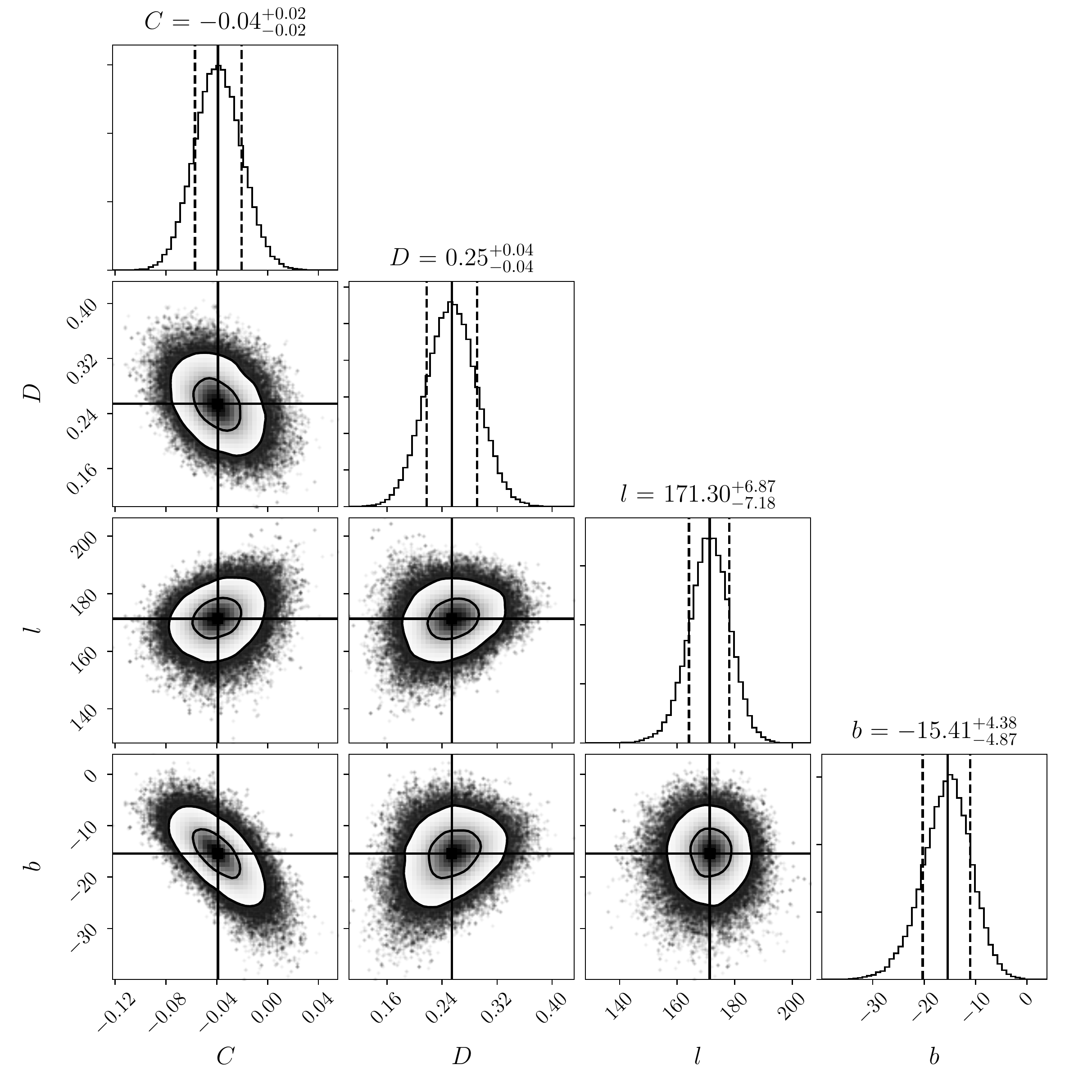}
\caption{The 1-dimensional marginalized histograms and 2-dimensional marginalized contours for the parameters space $\{C,D,l,b\}$. The illustration is same as Fig. \ref{fig:data}.}\label{fig:parameter}
\end{center}
\end{figure}

The MCMC result for parameter space $\{A,B,l,b\}$ is shown in Fig. \ref{fig:data}. For every parameter, its distribution is almost Gaussian and the slightly larger one of $1\sigma$ credible interval is regarded as its uncertainties. The $\hat{g}_{th}$-monopole term is $A=-0.01\pm0.01$. It is negligible. The $\hat{g}_{th}$-dipole magnitude is $B=0.10\pm0.01$. It is a significant signal for the anisotropy. And the $\hat{g}_{th}$-dipole direction points towards  $(l,b)=(176.54^{\circ}\pm7.55^{\circ},-16.59^{\circ}\pm5.11^{\circ})$. The MCMC result for another parameter space $\{C,D,l,b\}$ is shown in Fig. \ref{fig:parameter}. As the same as Fig. \ref{fig:data}, the distribution for each parameter is almost Gaussian. The $\hat{g}_{\dag}$-monopole term is $C=-0.04\pm0.02$. It is more significant than $\hat{g}_{th}$-monopole term, but it still be negligible. The $\hat{g}_{\dag}$-dipole magnitude is $D=0.25\pm0.04$. It is another significant signal for the anisotropy. The $\hat{g}_{\dag}$-dipole term directly indicates that the acceleration scale could be spatial variable. The $\hat{g}_{\dag}$-dipole direction points towards $(l,b)=(171.30^{\circ}\pm7.18^{\circ},-15.41^{\circ}\pm4.87^{\circ})$.

\begin{figure*}
\begin{center}
\includegraphics[width=\textwidth]{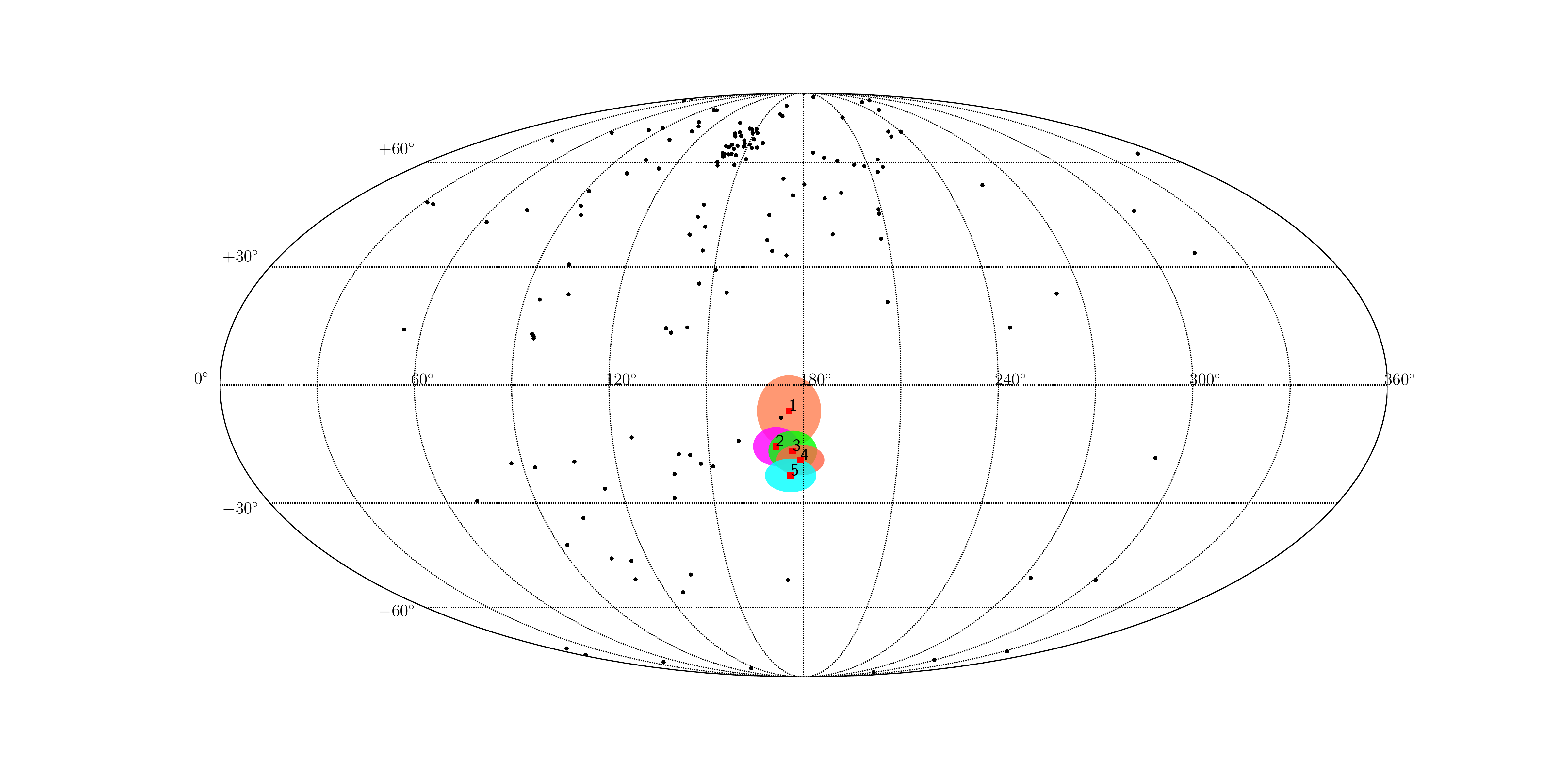}
\caption{The distribution of 147 SPARC galaxies on the sky (galactic coordinates system). Each point represents a single galaxy. The square point with its confidence region is labeled with number, which represents the direction of the hemisphere anisotropy ($2\sigma$) or dipole anisotropy ($1\sigma$). Specifically, they are: 1. the hemisphere anisotropy; 2. the $\hat{g}_{\dag}$-dipole anisotropy with monopole and dipole correction; 3. the $\hat{g}_{th}$-dipole anisotropy with monopole and dipole correction; 4. the $\hat{g}_{th}$-dipole anisotropy with dipole correction; 5. the $\hat{g}_{\dag}$-dipole anisotropy with dipole correction.}\label{fig:Map}
\end{center}
\end{figure*}

It's worth noting that the 2-dimensional marginalized contours for both parameter spaces have very similar shape. In addition, two dipole directions are very close to each other, and the angular separation is only $5.17^{\circ}$ (see Fig. \ref{fig:Map}). Furthermore, the minimal chi-square of the $\hat{g}_{\dag}$-dipole model is close to that of the $\hat{g}_{th}$-dipole model (see Table \ref{tab:result}). These results mean that the dipole anisotropy on the radial acceleration is mostly induced by the dipole anisotropy on acceleration scale. In our previous work \citep{Zhou:2017lwy}, we employed the hemisphere comparison method with the same SPARC data set to search possible spatial anisotropy on the acceleration scale. We found the maximum anisotropy direction is pointing to the direction $(l,b) = ({175.5^\circ}^{+6^\circ}_{-10^\circ}, {-6.5^\circ}^{+9^\circ}_{-3^\circ})$, which is very close to the $\hat{g}_{\dag}$-dipole direction and the angular separation is only $9.82^{\circ}$ (see Fig. \ref{fig:Map}).

For both the dipole models, the monopole term is indeed negligible comparing to the dipole magnitude, so we can neglect it from both corrections. Table \ref{tab:result} is the MCMC result for all dipole models with or without the monopole term.  Without the monopole term, both dipole magnitudes become slightly smaller, and the dipole directions are both shift a little to southeast with an angle less than $8.52^\circ$ (see Fig. \ref{fig:Map}). In addition, the chi-squares also have a slightly increase. These results indicate that the monopole term only has slight impact on the dipole anisotropy. Both AIC and BIC indicate that there is `decisive' evidence for all dipole models against the reference model, but it is indistinguishable to the dipole model with or without the monopole term. Even though there is `strong' evidence for $\hat{g}_{th}$-dipole model against $\hat{g}_{\dag}$-dipole model, these models could be compatible when the dipole anisotropy on radial acceleration is induced by the dipole anisotropy on acceleration scale.

\begin{table*}
\caption{The best fitting values with their $1\sigma$ uncertainties for four dipole models. The $\Delta\textrm{IC}$ value which is against the  reference model \eqref{eq:gth}, and the unnormalized $\chi^2$ for each correction are also listed at last three rows.}\label{tab:result}
\begin{tabular}{cccccccc}
  \hline
  Model  & $A(C)$ & $B(D)$ & $l$ & $b$ & $\chi^2$ & $\Delta\textrm{AIC}$ & $\Delta\textrm{BIC}$ \\
  \hline
  $\hat{g}_{th}$-dipole       & $-0.01\pm0.01$ & $0.10\pm0.01$    & $176.54^{\circ}\pm7.55^{\circ}$ & $-16.59^{\circ}\pm5.11^{\circ}$ & 3955  & -59 & -41 \\
                              &       -        & $0.09\pm0.01$    & $178.96^{\circ}\pm7.52^{\circ}$ & $-18.84^{\circ}\pm3.90^{\circ}$ & 3956  & -60 & -48 \\
  $\hat{g}_{\dag}$-dipole     & $-0.04\pm0.02$ & $0.25\pm0.04$    & $171.30^{\circ}\pm7.18^{\circ}$ & $-15.41^{\circ}\pm4.87^{\circ}$ & 3962  & -52 & -34 \\
                              &       -        & $0.23\pm0.04$    & $175.79^{\circ}\pm8.14^{\circ}$ & $-22.80^{\circ}\pm4.29^{\circ}$ & 3967  & -49 & -37 \\
  \hline
\end{tabular}
\end{table*}

\section{5. Monte Carlo simulations}
\label{Robust check}

As a robust check, we examine whether the dipole anisotropy could be derived from the statistical isotropy. First we create a mock data set from the SPARC data set. The dynamic centripetal acceleration $g_{obs}$ is replaced by a random number which has a Gaussian distribution, i.e. $G(g_{th},\sigma_{obs})$. Here $g_{th}$ is the theoretical centripetal acceleration \eqref{eq:gth} and the acceleration scale has been fixed at the best fitting value $g_{\dag}=\rm 1.02\times 10^{-10}~m~s^{-2}$. $\sigma_{obs}$ is the  uncertainty of $g_{obs}$. Except for the dynamic centripetal acceleration, other data including the galactic coordination, acceleration uncertainties and the baryonic acceleration remain unchanged. Then we employ the monopole and dipole correction \eqref{eq:data} and \eqref{eq:parameter} for the radial acceleration relation with the mock data set and use the MCMC method to explore the entire parameter space $\{A,B,l,b\}$ and $\{C,D,l,b\}$.

\begin{figure}
\begin{center}
\includegraphics[width=8.6cm]{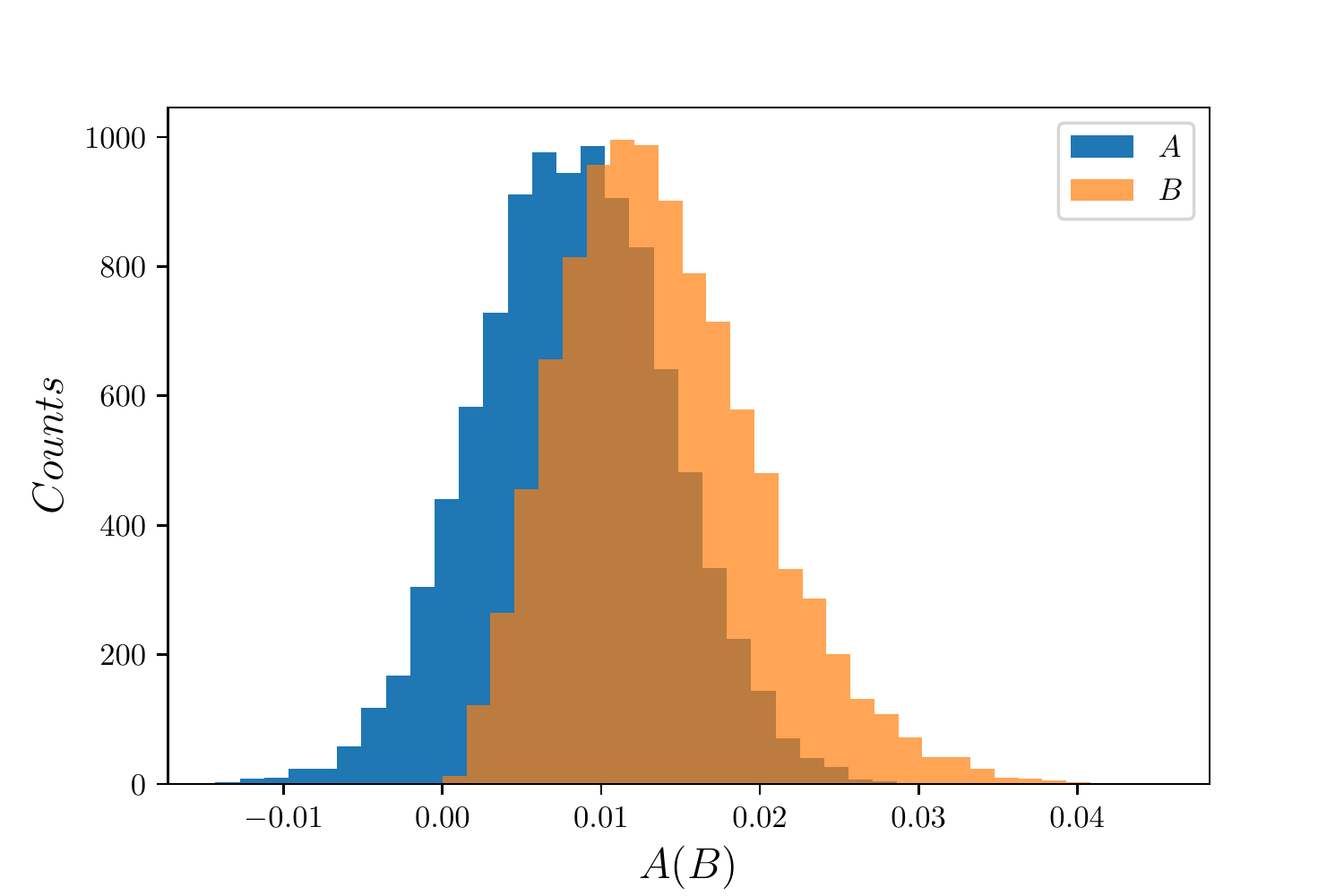}
\caption{The monopole term $A$ and the dipole magnitude $B$ in the isotropic mock data set for 10000 Monte Carlo simulations.}\label{fig:Isotropy1}
\end{center}
\end{figure}

\begin{figure}
\begin{center}
\includegraphics[width=8.6cm]{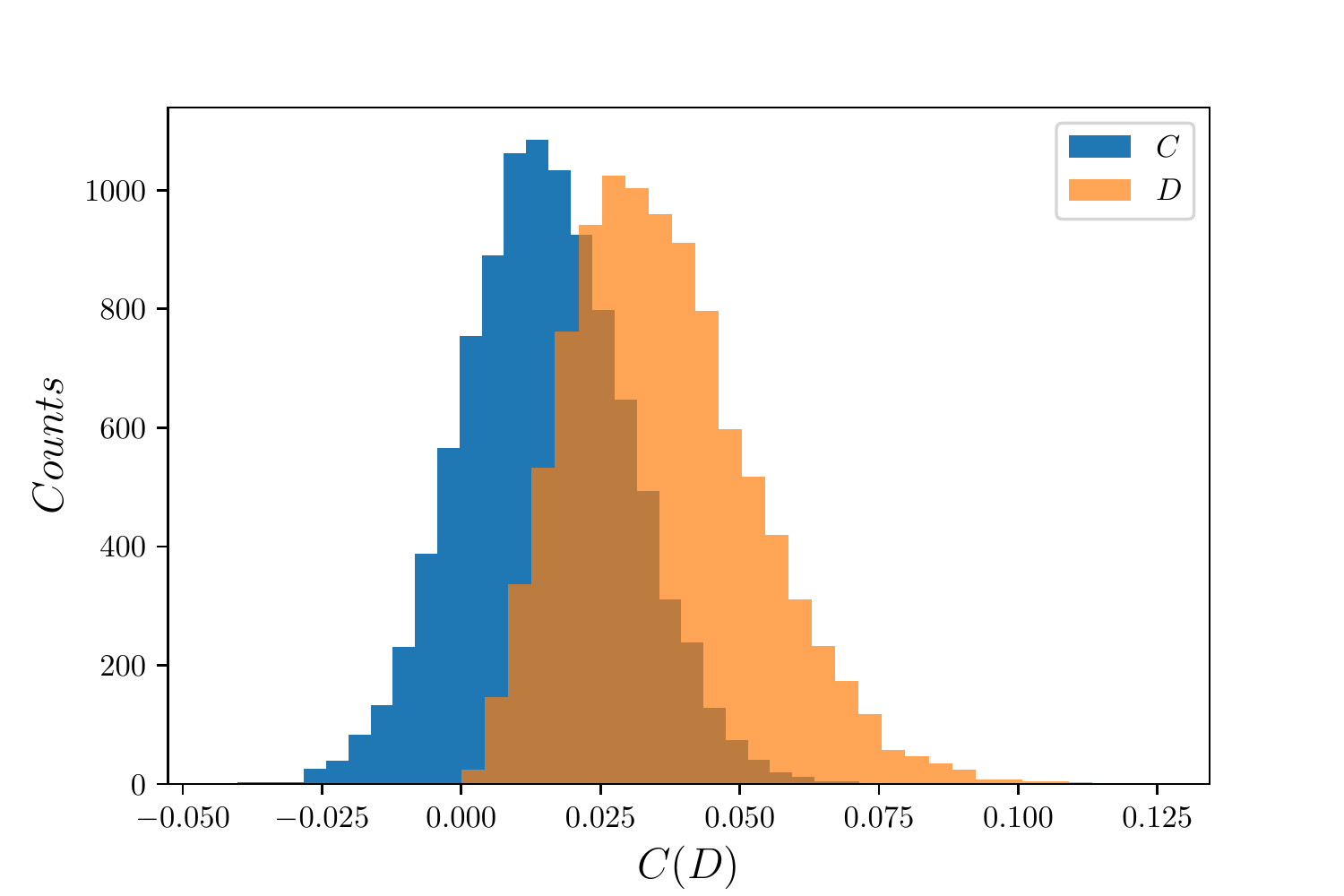}
\caption{The monopole term $C$ and the dipole magnitude $D$ in the isotropic mock data set for 10000 Monte Carlo simulations.}\label{fig:Isotropy2}
\end{center}
\end{figure}

For both the dipole models, we find that it is hard to constraint the dipole direction. It has a relatively large uncertainty and spans in all possible directions.  The monopole term is still negligible, but the dipole magnitude becomes much less than that from the SPARC data set. Fig. \ref{fig:Isotropy1} and \ref{fig:Isotropy2} are the results of 10000 simulations for the $\hat{g}_{th}$-dipole model and $\hat{g}_{\dag}$-dipole model (here we only use the ODR algorithm to fit the radial acceleration relation on account of the computation time). For the $\hat{g}_{th}$-dipole model, the monopole term centers on $\bar{A}=0.01$. It rarely reaches up to $A_{max}=0.03$, so that the monopole term still be negligible. The dipole magnitude centers on $\bar{B}=0.01$, and its upper limit only reaches up to $B_{max}=0.04$. It is much less than the dipole magnitude from the SPARC data set. For the $\hat{g}_{\dag}$-dipole model, the monopole term centers on $\bar{C}=0.01$. It rarely reaches up to $C_{max}=0.07$, so that the monopole term still be negligible. The dipole magnitude centers on $\bar{D}=0.03$, and its upper limit only reaches up to $D_{max}=0.10$. It is much less than the dipole magnitude from the SPARC data set. All these results mean that the dipole anisotropy from the original SPARC data set could not be reproduced by isotropic mock data set. This check is consistent with the robust check for the hemisphere comparison method \citep{Zhou:2017lwy}.

\section{6. Conclusions and discussions}
\label{Conclusions}

In this paper, we show a monopole and dipole correction for the radial acceleration relation with 147 rotationally supported galaxies. We found that there exist a significant dipole anisotropy on the radial acceleration, which is most probably induced by the dipole anisotropy on acceleration scale. The $\hat{g}_{\dag}$-dipole magnitude is significant and reaches up to $D =0.25\pm0.04$. The $\hat{g}_{\dag}$-dipole direction is pointing to the direction $(l,b) = (171.30^{\circ}\pm7.18^{\circ}, -15.41^{\circ}\pm4.87^{\circ})$, which is very close to the maximum anisotropy direction from the hemisphere comparison method. The monopole term is negligible. It only has slight impact on the dipole anisotropy. As the same as the hemisphere comparison method, the robust check has been taken to examine the significance of the dipole anisotropy. The result shows the dipole anisotropy couldn't be reproduced by the isotropic mock data set.

As pointed out at the introduction, the cosmological principle has been challenged by some cosmological observations. In this paper, we have found a possible dipole anisotropy on acceleration scale $g_{\dag}$ in local Universe. The dipole direction is very close to the cosmological preferred direction from the hemisphere comparison method \citep{Zhou:2017lwy}. These results hint that the Universe may be anisotropic and it could be related to some underlying physical effects, such as the spacetime anisotropy \citep{Chang:2011pza,Cahng:2013zwa,Li:2015uda}. If the cosmological principle is no longer valid, the standard $\Lambda$CDM model needs to be modified. However, it is still premature to claim that the Universe is anisotropic due to the small data samples and the uncertainty in the current observations.

There are some uncertainties in the original SPARC data set yet. As stated by \citet{McGaugh:2016leg}, the near-infrared (NIR) luminosity was observed while physics requires stellar (baryonic) mass. The mass-to-light ratio $\Upsilon_*$ is an unavoidable conversion factor which could be estimated by the stellar population synthesis (SPS) model \citep{2014PASA...31...11S}. The SPS model suggests that $\Upsilon_*$ is nearly constant in the NIR (within $\sim0.1$ dex), thus \citet{McGaugh:2016leg} assume a constant $\Upsilon_*$ for all galaxies to fit the radial acceleration relation. We use the same assumption in this paper as a precondition to search for the possible dipole ansitropy. Recently, \citet{Li:2018tdo} took $\Upsilon_*$ as `free' parameter to fit the radial acceleration relation to individual SPARC galaxies. If the possible small variation of $\Upsilon_*$ be taken into account, then the possible dipole anisotropy on acceleration scale may be impacted. Further investigations are necessary for seeking the possible degeneracy. Another possible uncertainty comes from the inhomogeneous distribution of galaxies on the sky (see Fig. \ref{fig:Map}). For future research on the anisotropy with galaxies, it is better to cover the sky homogeneously.

\begin{acknowledgments}
We are grateful to Dr. Yu Sang and Dong Zhao for useful discussions. We are also thankful for the open access of the SPARC data set. This work is supported by the National Natural Science Fund of China under grant Nos. 11675182, 11690022 and 11603005.
\end{acknowledgments}

\providecommand{\noopsort}[1]{}\providecommand{\singleletter}[1]{#1}%


\begin{thebibliography}{45}%
\makeatletter
\providecommand \@ifxundefined [1]{%
 \@ifx{#1\undefined}
}%
\providecommand \@ifnum [1]{%
 \ifnum #1\expandafter \@firstoftwo
 \else \expandafter \@secondoftwo
 \fi
}%
\providecommand \@ifx [1]{%
 \ifx #1\expandafter \@firstoftwo
 \else \expandafter \@secondoftwo
 \fi
}%
\providecommand \natexlab [1]{#1}%
\providecommand \enquote  [1]{``#1''}%
\providecommand \bibnamefont  [1]{#1}%
\providecommand \bibfnamefont [1]{#1}%
\providecommand \citenamefont [1]{#1}%
\providecommand \href@noop [0]{\@secondoftwo}%
\providecommand \href [0]{\begingroup \@sanitize@url \@href}%
\providecommand \@href[1]{\@@startlink{#1}\@@href}%
\providecommand \@@href[1]{\endgroup#1\@@endlink}%
\providecommand \@sanitize@url [0]{\catcode `\\12\catcode `\$12\catcode
  `\&12\catcode `\#12\catcode `\^12\catcode `\_12\catcode `\%12\relax}%
\providecommand \@@startlink[1]{}%
\providecommand \@@endlink[0]{}%
\providecommand \url  [0]{\begingroup\@sanitize@url \@url }%
\providecommand \@url [1]{\endgroup\@href {#1}{\urlprefix }}%
\providecommand \urlprefix  [0]{URL }%
\providecommand \Eprint [0]{\href }%
\providecommand \doibase [0]{http://dx.doi.org/}%
\providecommand \selectlanguage [0]{\@gobble}%
\providecommand \bibinfo  [0]{\@secondoftwo}%
\providecommand \bibfield  [0]{\@secondoftwo}%
\providecommand \translation [1]{[#1]}%
\providecommand \BibitemOpen [0]{}%
\providecommand \bibitemStop [0]{}%
\providecommand \bibitemNoStop [0]{.\EOS\space}%
\providecommand \EOS [0]{\spacefactor3000\relax}%
\providecommand \BibitemShut  [1]{\csname bibitem#1\endcsname}%
\let\auto@bib@innerbib\@empty
\bibitem [{\citenamefont {Weinberg}(2008)}]{Weinberg:2008zzc}%
  \BibitemOpen
  \bibfield  {author} {\bibinfo {author} {\bibfnamefont {S.}~\bibnamefont
  {Weinberg}},\ }\href
  {https://global.oup.com/academic/product/cosmology-9780198526827} {\emph
  {\bibinfo {title} {{Cosmology}}}}\ (\bibinfo {year} {2008})\BibitemShut
  {NoStop}%
\bibitem [{\citenamefont {Bennett}\ \emph {et~al.}(2013)\citenamefont {Bennett}
  \emph {et~al.}}]{Bennett:2012zja}%
  \BibitemOpen
  \bibfield  {author} {\bibinfo {author} {\bibfnamefont {C.~L.}\ \bibnamefont
  {Bennett}} \emph {et~al.} (\bibinfo {collaboration} {WMAP}),\ }\href
  {\doibase 10.1088/0067-0049/208/2/20} {\bibfield  {journal} {\bibinfo
  {journal} {Astrophys. J. Suppl.}\ }\textbf {\bibinfo {volume} {208}},\
  \bibinfo {pages} {20} (\bibinfo {year} {2013})},\ \Eprint
  {http://arxiv.org/abs/1212.5225} {arXiv:1212.5225 [astro-ph.CO]} \BibitemShut
  {NoStop}%
\bibitem [{\citenamefont {Hinshaw}\ \emph {et~al.}(2013)\citenamefont {Hinshaw}
  \emph {et~al.}}]{Hinshaw:2012aka}%
  \BibitemOpen
  \bibfield  {author} {\bibinfo {author} {\bibfnamefont {G.}~\bibnamefont
  {Hinshaw}} \emph {et~al.} (\bibinfo {collaboration} {WMAP}),\ }\href
  {\doibase 10.1088/0067-0049/208/2/19} {\bibfield  {journal} {\bibinfo
  {journal} {Astrophys. J. Suppl.}\ }\textbf {\bibinfo {volume} {208}},\
  \bibinfo {pages} {19} (\bibinfo {year} {2013})},\ \Eprint
  {http://arxiv.org/abs/1212.5226} {arXiv:1212.5226 [astro-ph.CO]} \BibitemShut
  {NoStop}%
\bibitem [{\citenamefont {Ade}\ \emph {et~al.}(2014)\citenamefont {Ade} \emph
  {et~al.}}]{Ade:2013zuv}%
  \BibitemOpen
  \bibfield  {author} {\bibinfo {author} {\bibfnamefont {P.~A.~R.}\
  \bibnamefont {Ade}} \emph {et~al.} (\bibinfo {collaboration} {Planck}),\
  }\href {\doibase 10.1051/0004-6361/201321591} {\bibfield  {journal} {\bibinfo
   {journal} {Astron. Astrophys.}\ }\textbf {\bibinfo {volume} {571}},\
  \bibinfo {pages} {A16} (\bibinfo {year} {2014})},\ \Eprint
  {http://arxiv.org/abs/1303.5076} {arXiv:1303.5076 [astro-ph.CO]} \BibitemShut
  {NoStop}%
\bibitem [{\citenamefont {Ade}\ \emph {et~al.}(2016)\citenamefont {Ade} \emph
  {et~al.}}]{Ade:2015xua}%
  \BibitemOpen
  \bibfield  {author} {\bibinfo {author} {\bibfnamefont {P.~A.~R.}\
  \bibnamefont {Ade}} \emph {et~al.} (\bibinfo {collaboration} {Planck}),\
  }\href {\doibase 10.1051/0004-6361/201525830} {\bibfield  {journal} {\bibinfo
   {journal} {Astron. Astrophys.}\ }\textbf {\bibinfo {volume} {594}},\
  \bibinfo {pages} {A13} (\bibinfo {year} {2016})},\ \Eprint
  {http://arxiv.org/abs/1502.01589} {arXiv:1502.01589 [astro-ph.CO]}
  \BibitemShut {NoStop}%
\bibitem [{\citenamefont {Hutsemekers}\ \emph {et~al.}(2005)\citenamefont
  {Hutsemekers}, \citenamefont {Cabanac}, \citenamefont {Lamy},\ and\
  \citenamefont {Sluse}}]{Hutsemekers:2005iz}%
  \BibitemOpen
  \bibfield  {author} {\bibinfo {author} {\bibfnamefont {D.}~\bibnamefont
  {Hutsemekers}}, \bibinfo {author} {\bibfnamefont {R.}~\bibnamefont
  {Cabanac}}, \bibinfo {author} {\bibfnamefont {H.}~\bibnamefont {Lamy}}, \
  and\ \bibinfo {author} {\bibfnamefont {D.}~\bibnamefont {Sluse}},\ }\href
  {\doibase 10.1051/0004-6361:20053337} {\bibfield  {journal} {\bibinfo
  {journal} {Astron. Astrophys.}\ }\textbf {\bibinfo {volume} {441}},\ \bibinfo
  {pages} {915} (\bibinfo {year} {2005})},\ \Eprint
  {http://arxiv.org/abs/astro-ph/0507274} {arXiv:astro-ph/0507274 [astro-ph]}
  \BibitemShut {NoStop}%
\bibitem [{\citenamefont {Kashlinsky}\ \emph {et~al.}(2010)\citenamefont
  {Kashlinsky}, \citenamefont {Atrio-Barandela}, \citenamefont {Ebeling},
  \citenamefont {Edge},\ and\ \citenamefont {Kocevski}}]{Kashlinsky:2009dw}%
  \BibitemOpen
  \bibfield  {author} {\bibinfo {author} {\bibfnamefont {A.}~\bibnamefont
  {Kashlinsky}}, \bibinfo {author} {\bibfnamefont {F.}~\bibnamefont
  {Atrio-Barandela}}, \bibinfo {author} {\bibfnamefont {H.}~\bibnamefont
  {Ebeling}}, \bibinfo {author} {\bibfnamefont {A.}~\bibnamefont {Edge}}, \
  and\ \bibinfo {author} {\bibfnamefont {D.}~\bibnamefont {Kocevski}},\ }\href
  {\doibase 10.1088/2041-8205/712/1/L81} {\bibfield  {journal} {\bibinfo
  {journal} {Astrophys. J.}\ }\textbf {\bibinfo {volume} {712}},\ \bibinfo
  {pages} {L81} (\bibinfo {year} {2010})},\ \Eprint
  {http://arxiv.org/abs/0910.4958} {arXiv:0910.4958 [astro-ph.CO]} \BibitemShut
  {NoStop}%
\bibitem [{\citenamefont {Feldman}\ \emph {et~al.}(2010)\citenamefont
  {Feldman}, \citenamefont {Watkins},\ and\ \citenamefont
  {Hudson}}]{Feldman:2009es}%
  \BibitemOpen
  \bibfield  {author} {\bibinfo {author} {\bibfnamefont {H.~A.}\ \bibnamefont
  {Feldman}}, \bibinfo {author} {\bibfnamefont {R.}~\bibnamefont {Watkins}}, \
  and\ \bibinfo {author} {\bibfnamefont {M.~J.}\ \bibnamefont {Hudson}},\
  }\href {\doibase 10.1111/j.1365-2966.2010.17052.x} {\bibfield  {journal}
  {\bibinfo  {journal} {Mon. Not. Roy. Astron. Soc.}\ }\textbf {\bibinfo
  {volume} {407}},\ \bibinfo {pages} {2328} (\bibinfo {year} {2010})},\ \Eprint
  {http://arxiv.org/abs/0911.5516} {arXiv:0911.5516 [astro-ph.CO]} \BibitemShut
  {NoStop}%
\bibitem [{\citenamefont {Webb}\ \emph {et~al.}(2011)\citenamefont {Webb},
  \citenamefont {King}, \citenamefont {Murphy}, \citenamefont {Flambaum},
  \citenamefont {Carswell},\ and\ \citenamefont {Bainbridge}}]{Webb:2010hc}%
  \BibitemOpen
  \bibfield  {author} {\bibinfo {author} {\bibfnamefont {J.~K.}\ \bibnamefont
  {Webb}}, \bibinfo {author} {\bibfnamefont {J.~A.}\ \bibnamefont {King}},
  \bibinfo {author} {\bibfnamefont {M.~T.}\ \bibnamefont {Murphy}}, \bibinfo
  {author} {\bibfnamefont {V.~V.}\ \bibnamefont {Flambaum}}, \bibinfo {author}
  {\bibfnamefont {R.~F.}\ \bibnamefont {Carswell}}, \ and\ \bibinfo {author}
  {\bibfnamefont {M.~B.}\ \bibnamefont {Bainbridge}},\ }\href {\doibase
  10.1103/PhysRevLett.107.191101} {\bibfield  {journal} {\bibinfo  {journal}
  {Phys. Rev. Lett.}\ }\textbf {\bibinfo {volume} {107}},\ \bibinfo {pages}
  {191101} (\bibinfo {year} {2011})},\ \Eprint {http://arxiv.org/abs/1008.3907}
  {arXiv:1008.3907 [astro-ph.CO]} \BibitemShut {NoStop}%
\bibitem [{\citenamefont {King}\ \emph {et~al.}(2012)\citenamefont {King},
  \citenamefont {Webb}, \citenamefont {Murphy}, \citenamefont {Flambaum},
  \citenamefont {Carswell}, \citenamefont {Bainbridge}, \citenamefont
  {Wilczynska},\ and\ \citenamefont {Koch}}]{King:2012id}%
  \BibitemOpen
  \bibfield  {author} {\bibinfo {author} {\bibfnamefont {J.~A.}\ \bibnamefont
  {King}}, \bibinfo {author} {\bibfnamefont {J.~K.}\ \bibnamefont {Webb}},
  \bibinfo {author} {\bibfnamefont {M.~T.}\ \bibnamefont {Murphy}}, \bibinfo
  {author} {\bibfnamefont {V.~V.}\ \bibnamefont {Flambaum}}, \bibinfo {author}
  {\bibfnamefont {R.~F.}\ \bibnamefont {Carswell}}, \bibinfo {author}
  {\bibfnamefont {M.~B.}\ \bibnamefont {Bainbridge}}, \bibinfo {author}
  {\bibfnamefont {M.~R.}\ \bibnamefont {Wilczynska}}, \ and\ \bibinfo {author}
  {\bibfnamefont {F.~E.}\ \bibnamefont {Koch}},\ }\href {\doibase
  10.1111/j.1365-2966.2012.20852.x} {\bibfield  {journal} {\bibinfo  {journal}
  {Mon. Not. Roy. Astron. Soc.}\ }\textbf {\bibinfo {volume} {422}},\ \bibinfo
  {pages} {3370} (\bibinfo {year} {2012})},\ \Eprint
  {http://arxiv.org/abs/1202.4758} {arXiv:1202.4758 [astro-ph.CO]} \BibitemShut
  {NoStop}%
\bibitem [{\citenamefont {Mariano}\ and\ \citenamefont
  {Perivolaropoulos}(2012)}]{Mariano:2012wx}%
  \BibitemOpen
  \bibfield  {author} {\bibinfo {author} {\bibfnamefont {A.}~\bibnamefont
  {Mariano}}\ and\ \bibinfo {author} {\bibfnamefont {L.}~\bibnamefont
  {Perivolaropoulos}},\ }\href {\doibase 10.1103/PhysRevD.86.083517} {\bibfield
   {journal} {\bibinfo  {journal} {Phys. Rev.}\ }\textbf {\bibinfo {volume}
  {D86}},\ \bibinfo {pages} {083517} (\bibinfo {year} {2012})},\ \Eprint
  {http://arxiv.org/abs/1206.4055} {arXiv:1206.4055 [astro-ph.CO]} \BibitemShut
  {NoStop}%
\bibitem [{\citenamefont {Chang}\ \emph {et~al.}(2014)\citenamefont {Chang},
  \citenamefont {Li}, \citenamefont {Lin},\ and\ \citenamefont
  {Wang}}]{Chang:2014wpa}%
  \BibitemOpen
  \bibfield  {author} {\bibinfo {author} {\bibfnamefont {Z.}~\bibnamefont
  {Chang}}, \bibinfo {author} {\bibfnamefont {X.}~\bibnamefont {Li}}, \bibinfo
  {author} {\bibfnamefont {H.-N.}\ \bibnamefont {Lin}}, \ and\ \bibinfo
  {author} {\bibfnamefont {S.}~\bibnamefont {Wang}},\ }\href {\doibase
  10.1140/epjc/s10052-014-2821-7} {\bibfield  {journal} {\bibinfo  {journal}
  {Eur. Phys. J.}\ }\textbf {\bibinfo {volume} {C74}},\ \bibinfo {pages} {2821}
  (\bibinfo {year} {2014})},\ \Eprint {http://arxiv.org/abs/1403.5661}
  {arXiv:1403.5661 [astro-ph.CO]} \BibitemShut {NoStop}%
\bibitem [{\citenamefont {Chang}\ and\ \citenamefont
  {Lin}(2015)}]{Chang:2014nca}%
  \BibitemOpen
  \bibfield  {author} {\bibinfo {author} {\bibfnamefont {Z.}~\bibnamefont
  {Chang}}\ and\ \bibinfo {author} {\bibfnamefont {H.-N.}\ \bibnamefont
  {Lin}},\ }\href {\doibase 10.1093/mnras/stu2349} {\bibfield  {journal}
  {\bibinfo  {journal} {Mon. Not. Roy. Astron. Soc.}\ }\textbf {\bibinfo
  {volume} {446}},\ \bibinfo {pages} {2952} (\bibinfo {year} {2015})},\ \Eprint
  {http://arxiv.org/abs/1411.1466} {arXiv:1411.1466 [astro-ph.CO]} \BibitemShut
  {NoStop}%
\bibitem [{\citenamefont {Lin}\ \emph {et~al.}(2016)\citenamefont {Lin},
  \citenamefont {Li},\ and\ \citenamefont {Chang}}]{Lin:2016jqp}%
  \BibitemOpen
  \bibfield  {author} {\bibinfo {author} {\bibfnamefont {H.-N.}\ \bibnamefont
  {Lin}}, \bibinfo {author} {\bibfnamefont {X.}~\bibnamefont {Li}}, \ and\
  \bibinfo {author} {\bibfnamefont {Z.}~\bibnamefont {Chang}},\ }\href
  {\doibase 10.1093/mnras/stw995} {\bibfield  {journal} {\bibinfo  {journal}
  {Mon. Not. Roy. Astron. Soc.}\ }\textbf {\bibinfo {volume} {460}},\ \bibinfo
  {pages} {617} (\bibinfo {year} {2016})},\ \Eprint
  {http://arxiv.org/abs/1604.07505} {arXiv:1604.07505 [astro-ph.CO]}
  \BibitemShut {NoStop}%
\bibitem [{\citenamefont {Rubin}\ \emph {et~al.}(1978)\citenamefont {Rubin},
  \citenamefont {Ford},\ and\ \citenamefont {Thonnard}}]{Rubin:1978kmz}%
  \BibitemOpen
  \bibfield  {author} {\bibinfo {author} {\bibfnamefont {V.~C.}\ \bibnamefont
  {Rubin}}, \bibinfo {author} {\bibfnamefont {W.~K.}\ \bibnamefont {Ford},
  \bibfnamefont {Jr.}}, \ and\ \bibinfo {author} {\bibfnamefont
  {N.}~\bibnamefont {Thonnard}},\ }\href {\doibase 10.1086/182804} {\bibfield
  {journal} {\bibinfo  {journal} {Astrophys. J.}\ }\textbf {\bibinfo {volume}
  {225}},\ \bibinfo {pages} {L107} (\bibinfo {year} {1978})}\BibitemShut
  {NoStop}%
\bibitem [{\citenamefont {Bosma}(1981)}]{Bosma:1981zz}%
  \BibitemOpen
  \bibfield  {author} {\bibinfo {author} {\bibfnamefont {A.}~\bibnamefont
  {Bosma}},\ }\href {\doibase 10.1086/113063} {\bibfield  {journal} {\bibinfo
  {journal} {Astron. J.}\ }\textbf {\bibinfo {volume} {86}},\ \bibinfo {pages}
  {1825} (\bibinfo {year} {1981})}\BibitemShut {NoStop}%
\bibitem [{\citenamefont {Tan}\ \emph {et~al.}(2016)\citenamefont {Tan} \emph
  {et~al.}}]{Tan:2016zwf}%
  \BibitemOpen
  \bibfield  {author} {\bibinfo {author} {\bibfnamefont {A.}~\bibnamefont
  {Tan}} \emph {et~al.} (\bibinfo {collaboration} {PandaX-II}),\ }\href
  {\doibase 10.1103/PhysRevLett.117.121303} {\bibfield  {journal} {\bibinfo
  {journal} {Phys. Rev. Lett.}\ }\textbf {\bibinfo {volume} {117}},\ \bibinfo
  {pages} {121303} (\bibinfo {year} {2016})},\ \Eprint
  {http://arxiv.org/abs/1607.07400} {arXiv:1607.07400 [hep-ex]} \BibitemShut
  {NoStop}%
\bibitem [{\citenamefont {Aprile}\ \emph {et~al.}(2016)\citenamefont {Aprile}
  \emph {et~al.}}]{Aprile:2016swn}%
  \BibitemOpen
  \bibfield  {author} {\bibinfo {author} {\bibfnamefont {E.}~\bibnamefont
  {Aprile}} \emph {et~al.} (\bibinfo {collaboration} {XENON100}),\ }\href
  {\doibase 10.1103/PhysRevD.94.122001} {\bibfield  {journal} {\bibinfo
  {journal} {Phys. Rev.}\ }\textbf {\bibinfo {volume} {D94}},\ \bibinfo {pages}
  {122001} (\bibinfo {year} {2016})},\ \Eprint
  {http://arxiv.org/abs/1609.06154} {arXiv:1609.06154 [astro-ph.CO]}
  \BibitemShut {NoStop}%
\bibitem [{\citenamefont {Milgrom}(1983)}]{Milgrom:1983ca}%
  \BibitemOpen
  \bibfield  {author} {\bibinfo {author} {\bibfnamefont {M.}~\bibnamefont
  {Milgrom}},\ }\href {\doibase 10.1086/161130} {\bibfield  {journal} {\bibinfo
   {journal} {Astrophys. J.}\ }\textbf {\bibinfo {volume} {270}},\ \bibinfo
  {pages} {365} (\bibinfo {year} {1983})}\BibitemShut {NoStop}%
\bibitem [{\citenamefont {Milgrom}(2002)}]{Milgrom:2002tu}%
  \BibitemOpen
  \bibfield  {author} {\bibinfo {author} {\bibfnamefont {M.}~\bibnamefont
  {Milgrom}},\ }\href {\doibase 10.1016/S1387-6473(02)00243-9} {\bibfield
  {journal} {\bibinfo  {journal} {New Astron. Rev.}\ }\textbf {\bibinfo
  {volume} {46}},\ \bibinfo {pages} {741} (\bibinfo {year} {2002})},\ \Eprint
  {http://arxiv.org/abs/astro-ph/0207231} {arXiv:astro-ph/0207231 [astro-ph]}
  \BibitemShut {NoStop}%
\bibitem [{\citenamefont {Milgrom}(2014)}]{Milgrom:2012xw}%
  \BibitemOpen
  \bibfield  {author} {\bibinfo {author} {\bibfnamefont {M.}~\bibnamefont
  {Milgrom}},\ }\href {\doibase 10.1093/mnras/stt2066} {\bibfield  {journal}
  {\bibinfo  {journal} {Mon. Not. Roy. Astron. Soc.}\ }\textbf {\bibinfo
  {volume} {437}},\ \bibinfo {pages} {2531} (\bibinfo {year} {2014})},\ \Eprint
  {http://arxiv.org/abs/1212.2568} {arXiv:1212.2568 [astro-ph.CO]} \BibitemShut
  {NoStop}%
\bibitem [{\citenamefont {Begeman}\ \emph {et~al.}(1991)\citenamefont
  {Begeman}, \citenamefont {Broeils},\ and\ \citenamefont
  {Sanders}}]{Begeman:1991iy}%
  \BibitemOpen
  \bibfield  {author} {\bibinfo {author} {\bibfnamefont {K.~G.}\ \bibnamefont
  {Begeman}}, \bibinfo {author} {\bibfnamefont {A.~H.}\ \bibnamefont
  {Broeils}}, \ and\ \bibinfo {author} {\bibfnamefont {R.~H.}\ \bibnamefont
  {Sanders}},\ }\href {\doibase 10.1093/mnras/249.3.523} {\bibfield  {journal}
  {\bibinfo  {journal} {Mon. Not. Roy. Astron. Soc.}\ }\textbf {\bibinfo
  {volume} {249}},\ \bibinfo {pages} {523} (\bibinfo {year}
  {1991})}\BibitemShut {NoStop}%
\bibitem [{\citenamefont {Swaters}\ \emph {et~al.}(2010)\citenamefont
  {Swaters}, \citenamefont {Sanders},\ and\ \citenamefont
  {McGaugh}}]{Swaters:2010qe}%
  \BibitemOpen
  \bibfield  {author} {\bibinfo {author} {\bibfnamefont {R.~A.}\ \bibnamefont
  {Swaters}}, \bibinfo {author} {\bibfnamefont {R.~H.}\ \bibnamefont
  {Sanders}}, \ and\ \bibinfo {author} {\bibfnamefont {S.~S.}\ \bibnamefont
  {McGaugh}},\ }\href {\doibase 10.1088/0004-637X/718/1/380} {\bibfield
  {journal} {\bibinfo  {journal} {Astrophys. J.}\ }\textbf {\bibinfo {volume}
  {718}},\ \bibinfo {pages} {380} (\bibinfo {year} {2010})},\ \Eprint
  {http://arxiv.org/abs/1005.5456} {arXiv:1005.5456 [astro-ph.CO]} \BibitemShut
  {NoStop}%
\bibitem [{\citenamefont {Chang}\ \emph
  {et~al.}(2013{\natexlab{a}})\citenamefont {Chang}, \citenamefont {Li},
  \citenamefont {Li}, \citenamefont {Lin},\ and\ \citenamefont
  {Wang}}]{Chang:2013twa}%
  \BibitemOpen
  \bibfield  {author} {\bibinfo {author} {\bibfnamefont {Z.}~\bibnamefont
  {Chang}}, \bibinfo {author} {\bibfnamefont {M.-H.}\ \bibnamefont {Li}},
  \bibinfo {author} {\bibfnamefont {X.}~\bibnamefont {Li}}, \bibinfo {author}
  {\bibfnamefont {H.-N.}\ \bibnamefont {Lin}}, \ and\ \bibinfo {author}
  {\bibfnamefont {S.}~\bibnamefont {Wang}},\ }\href {\doibase
  10.1140/epjc/s10052-013-2447-1} {\bibfield  {journal} {\bibinfo  {journal}
  {Eur. Phys. J.}\ }\textbf {\bibinfo {volume} {C73}},\ \bibinfo {pages} {2447}
  (\bibinfo {year} {2013}{\natexlab{a}})},\ \Eprint
  {http://arxiv.org/abs/1305.2314} {arXiv:1305.2314 [astro-ph.GA]} \BibitemShut
  {NoStop}%
\bibitem [{\citenamefont {Milgrom}(1998)}]{Milgrom:1998aj}%
  \BibitemOpen
  \bibfield  {author} {\bibinfo {author} {\bibfnamefont {M.}~\bibnamefont
  {Milgrom}},\ }in\ \href
  {https://inspirehep.net/record/483371/files/arXiv:astro-ph_9810302.pdf}
  {\emph {\bibinfo {booktitle} {{Proceedings, 2nd International Heidelberg
  Conference on Dark matter in astrophysics and particle physics (DARK 1998):
  Heidelberg, Germany, July 20-25, 1998}}}}\ (\bibinfo {year} {1998})\ pp.\
  \bibinfo {pages} {443--457},\ \Eprint {http://arxiv.org/abs/astro-ph/9810302}
  {arXiv:astro-ph/9810302 [astro-ph]} \BibitemShut {NoStop}%
\bibitem [{\citenamefont {Zhou}\ \emph {et~al.}(2017)\citenamefont {Zhou},
  \citenamefont {Zhao},\ and\ \citenamefont {Chang}}]{Zhou:2017lwy}%
  \BibitemOpen
  \bibfield  {author} {\bibinfo {author} {\bibfnamefont {Y.}~\bibnamefont
  {Zhou}}, \bibinfo {author} {\bibfnamefont {Z.-C.}\ \bibnamefont {Zhao}}, \
  and\ \bibinfo {author} {\bibfnamefont {Z.}~\bibnamefont {Chang}},\ }\href
  {\doibase 10.3847/1538-4357/aa8991} {\bibfield  {journal} {\bibinfo
  {journal} {Astrophys. J.}\ }\textbf {\bibinfo {volume} {847}},\ \bibinfo
  {pages} {86} (\bibinfo {year} {2017})},\ \Eprint
  {http://arxiv.org/abs/1707.00417} {arXiv:1707.00417 [astro-ph.CO]}
  \BibitemShut {NoStop}%
\bibitem [{\citenamefont {{McGaugh}}\ \emph {et~al.}(2016)\citenamefont
  {{McGaugh}}, \citenamefont {{Lelli}},\ and\ \citenamefont
  {{Schombert}}}]{McGaugh:2016leg}%
  \BibitemOpen
  \bibfield  {author} {\bibinfo {author} {\bibfnamefont {S.~S.}\ \bibnamefont
  {{McGaugh}}}, \bibinfo {author} {\bibfnamefont {F.}~\bibnamefont {{Lelli}}},
  \ and\ \bibinfo {author} {\bibfnamefont {J.~M.}\ \bibnamefont
  {{Schombert}}},\ }\href {\doibase 10.1103/PhysRevLett.117.201101} {\bibfield
  {journal} {\bibinfo  {journal} {Physical Review Letters}\ }\textbf {\bibinfo
  {volume} {117}},\ \bibinfo {eid} {201101} (\bibinfo {year} {2016})},\ \Eprint
  {http://arxiv.org/abs/1609.05917} {arXiv:1609.05917} \BibitemShut {NoStop}%
\bibitem [{Note1()}]{Note1}%
  \BibitemOpen
  \bibinfo {note} {\protect \url {http://astroweb.cwru.edu/SPARC/}}\BibitemShut
  {NoStop}%
\bibitem [{\citenamefont {Lelli}\ \emph {et~al.}(2016)\citenamefont {Lelli},
  \citenamefont {McGaugh},\ and\ \citenamefont {Schombert}}]{Lelli:2016zqa}%
  \BibitemOpen
  \bibfield  {author} {\bibinfo {author} {\bibfnamefont {F.}~\bibnamefont
  {Lelli}}, \bibinfo {author} {\bibfnamefont {S.~S.}\ \bibnamefont {McGaugh}},
  \ and\ \bibinfo {author} {\bibfnamefont {J.~M.}\ \bibnamefont {Schombert}},\
  }\href {\doibase 10.3847/0004-6256/152/6/157} {\bibfield  {journal} {\bibinfo
   {journal} {Astron. J.}\ }\textbf {\bibinfo {volume} {152}},\ \bibinfo
  {pages} {157} (\bibinfo {year} {2016})},\ \Eprint
  {http://arxiv.org/abs/1606.09251} {arXiv:1606.09251 [astro-ph.GA]}
  \BibitemShut {NoStop}%
\bibitem [{\citenamefont {Begum}\ and\ \citenamefont
  {Chengalur}(2005)}]{Begum:2004sj}%
  \BibitemOpen
  \bibfield  {author} {\bibinfo {author} {\bibfnamefont {A.}~\bibnamefont
  {Begum}}\ and\ \bibinfo {author} {\bibfnamefont {J.~N.}\ \bibnamefont
  {Chengalur}},\ }\href {\doibase 10.1051/0004-6361:20041210} {\bibfield
  {journal} {\bibinfo  {journal} {Astron. Astrophys.}\ }\textbf {\bibinfo
  {volume} {424}},\ \bibinfo {pages} {509} (\bibinfo {year} {2005})},\ \Eprint
  {http://arxiv.org/abs/astro-ph/0406211} {arXiv:astro-ph/0406211 [astro-ph]}
  \BibitemShut {NoStop}%
\bibitem [{\citenamefont {de~Blok}\ \emph {et~al.}(1996)\citenamefont
  {de~Blok}, \citenamefont {McGaugh},\ and\ \citenamefont {van~der
  Hulst}}]{deBlok:1996jib}%
  \BibitemOpen
  \bibfield  {author} {\bibinfo {author} {\bibfnamefont {W.~J.~G.}\
  \bibnamefont {de~Blok}}, \bibinfo {author} {\bibfnamefont {S.~S.}\
  \bibnamefont {McGaugh}}, \ and\ \bibinfo {author} {\bibfnamefont {J.~M.}\
  \bibnamefont {van~der Hulst}},\ }\href {\doibase 10.1093/mnras/283.1.18}
  {\bibfield  {journal} {\bibinfo  {journal} {Mon. Not. Roy. Astron. Soc.}\
  }\textbf {\bibinfo {volume} {283}},\ \bibinfo {pages} {18} (\bibinfo {year}
  {1996})},\ \Eprint {http://arxiv.org/abs/astro-ph/9605069}
  {arXiv:astro-ph/9605069 [astro-ph]} \BibitemShut {NoStop}%
\bibitem [{Note2()}]{Note2}%
  \BibitemOpen
  \bibinfo {note} {\protect \url {http://ned.ipac.caltech.edu/}}\BibitemShut
  {NoStop}%
\bibitem [{\citenamefont {Milgrom}(2016)}]{Milgrom:2016uye}%
  \BibitemOpen
  \bibfield  {author} {\bibinfo {author} {\bibfnamefont {M.}~\bibnamefont
  {Milgrom}},\ }\href@noop {} {\  (\bibinfo {year} {2016})},\ \Eprint
  {http://arxiv.org/abs/1609.06642} {arXiv:1609.06642 [astro-ph.GA]}
  \BibitemShut {NoStop}%
\bibitem [{\citenamefont {Boggs}\ \emph {et~al.}(1987)\citenamefont {Boggs},
  \citenamefont {Byrd},\ and\ \citenamefont {Schnabel}}]{boggs1987stable}%
  \BibitemOpen
  \bibfield  {author} {\bibinfo {author} {\bibfnamefont {P.~T.}\ \bibnamefont
  {Boggs}}, \bibinfo {author} {\bibfnamefont {R.~H.}\ \bibnamefont {Byrd}}, \
  and\ \bibinfo {author} {\bibfnamefont {R.~B.}\ \bibnamefont {Schnabel}},\
  }\href {\doibase 10.1137/0908085} {\bibfield  {journal} {\bibinfo  {journal}
  {SIAM Journal on Scientific and Statistical Computing}\ }\textbf {\bibinfo
  {volume} {8}},\ \bibinfo {pages} {1052} (\bibinfo {year} {1987})}\BibitemShut
  {NoStop}%
\bibitem [{\citenamefont {{Akaike}}(1974)}]{1974ITAC...19..716A}%
  \BibitemOpen
  \bibfield  {author} {\bibinfo {author} {\bibfnamefont {H.}~\bibnamefont
  {{Akaike}}},\ }\href {\doibase 10.1109/TAC.1974.1100705} {\bibfield
  {journal} {\bibinfo  {journal} {IEEE Transactions on Automatic Control}\
  }\textbf {\bibinfo {volume} {19}},\ \bibinfo {pages} {716} (\bibinfo {year}
  {1974})}\BibitemShut {NoStop}%
\bibitem [{\citenamefont {Schwarz}(1978)}]{1978AnSta...6..461S}%
  \BibitemOpen
  \bibfield  {author} {\bibinfo {author} {\bibfnamefont {G.}~\bibnamefont
  {Schwarz}},\ }\href {\doibase 10.1214/aos/1176344136} {\bibfield  {journal}
  {\bibinfo  {journal} {Annals of Statistics}\ }\textbf {\bibinfo {volume}
  {6}},\ \bibinfo {pages} {461} (\bibinfo {year} {1978})}\BibitemShut {NoStop}%
\bibitem [{\citenamefont {Liddle}(2007)}]{Liddle:2007fy}%
  \BibitemOpen
  \bibfield  {author} {\bibinfo {author} {\bibfnamefont {A.~R.}\ \bibnamefont
  {Liddle}},\ }\href {\doibase 10.1111/j.1745-3933.2007.00306.x} {\bibfield
  {journal} {\bibinfo  {journal} {Mon. Not. Roy. Astron. Soc.}\ }\textbf
  {\bibinfo {volume} {377}},\ \bibinfo {pages} {L74} (\bibinfo {year}
  {2007})},\ \Eprint {http://arxiv.org/abs/astro-ph/0701113}
  {arXiv:astro-ph/0701113 [astro-ph]} \BibitemShut {NoStop}%
\bibitem [{\citenamefont {Arevalo}\ \emph {et~al.}(2017)\citenamefont
  {Arevalo}, \citenamefont {Cid},\ and\ \citenamefont
  {Moya}}]{Arevalo:2016epc}%
  \BibitemOpen
  \bibfield  {author} {\bibinfo {author} {\bibfnamefont {F.}~\bibnamefont
  {Arevalo}}, \bibinfo {author} {\bibfnamefont {A.}~\bibnamefont {Cid}}, \ and\
  \bibinfo {author} {\bibfnamefont {J.}~\bibnamefont {Moya}},\ }\href {\doibase
  10.1140/epjc/s10052-017-5128-7} {\bibfield  {journal} {\bibinfo  {journal}
  {Eur. Phys. J.}\ }\textbf {\bibinfo {volume} {C77}},\ \bibinfo {pages} {565}
  (\bibinfo {year} {2017})},\ \Eprint {http://arxiv.org/abs/1610.09330}
  {arXiv:1610.09330 [astro-ph.CO]} \BibitemShut {NoStop}%
\bibitem [{\citenamefont {Lin}\ \emph {et~al.}(2017)\citenamefont {Lin},
  \citenamefont {Li},\ and\ \citenamefont {Sang}}]{Lin:2017yfl}%
  \BibitemOpen
  \bibfield  {author} {\bibinfo {author} {\bibfnamefont {H.-N.}\ \bibnamefont
  {Lin}}, \bibinfo {author} {\bibfnamefont {X.}~\bibnamefont {Li}}, \ and\
  \bibinfo {author} {\bibfnamefont {Y.}~\bibnamefont {Sang}},\ }\href@noop {}
  {\  (\bibinfo {year} {2017})},\ \Eprint {http://arxiv.org/abs/1711.05025}
  {arXiv:1711.05025 [astro-ph.CO]} \BibitemShut {NoStop}%
\bibitem [{\citenamefont {Foreman-Mackey}\ \emph {et~al.}(2013)\citenamefont
  {Foreman-Mackey}, \citenamefont {Hogg}, \citenamefont {Lang},\ and\
  \citenamefont {Goodman}}]{2013PASP..125..306F}%
  \BibitemOpen
  \bibfield  {author} {\bibinfo {author} {\bibfnamefont {D.}~\bibnamefont
  {Foreman-Mackey}}, \bibinfo {author} {\bibfnamefont {D.~W.}\ \bibnamefont
  {Hogg}}, \bibinfo {author} {\bibfnamefont {D.}~\bibnamefont {Lang}}, \ and\
  \bibinfo {author} {\bibfnamefont {J.}~\bibnamefont {Goodman}},\ }\href
  {\doibase 10.1086/670067} {\bibfield  {journal} {\bibinfo  {journal}
  {Publications of the Astronomical Society of the Pacific}\ }\textbf {\bibinfo
  {volume} {125}},\ \bibinfo {pages} {306} (\bibinfo {year} {2013})},\ \Eprint
  {http://arxiv.org/abs/1202.3665} {arXiv:1202.3665 [astro-ph.IM]} \BibitemShut
  {NoStop}%
\bibitem [{\citenamefont {Chang}\ \emph {et~al.}(2012)\citenamefont {Chang},
  \citenamefont {Wang},\ and\ \citenamefont {Li}}]{Chang:2011pza}%
  \BibitemOpen
  \bibfield  {author} {\bibinfo {author} {\bibfnamefont {Z.}~\bibnamefont
  {Chang}}, \bibinfo {author} {\bibfnamefont {S.}~\bibnamefont {Wang}}, \ and\
  \bibinfo {author} {\bibfnamefont {X.}~\bibnamefont {Li}},\ }\href {\doibase
  10.1140/epjc/s10052-011-1838-4} {\bibfield  {journal} {\bibinfo  {journal}
  {Eur. Phys. J.}\ }\textbf {\bibinfo {volume} {C72}},\ \bibinfo {pages} {1838}
  (\bibinfo {year} {2012})},\ \Eprint {http://arxiv.org/abs/1106.2726}
  {arXiv:1106.2726 [gr-qc]} \BibitemShut {NoStop}%
\bibitem [{\citenamefont {Chang}\ \emph
  {et~al.}(2013{\natexlab{b}})\citenamefont {Chang}, \citenamefont {Li},\ and\
  \citenamefont {Wang}}]{Cahng:2013zwa}%
  \BibitemOpen
  \bibfield  {author} {\bibinfo {author} {\bibfnamefont {Z.}~\bibnamefont
  {Chang}}, \bibinfo {author} {\bibfnamefont {M.-H.}\ \bibnamefont {Li}}, \
  and\ \bibinfo {author} {\bibfnamefont {S.}~\bibnamefont {Wang}},\ }\href
  {\doibase 10.1016/j.physletb.2013.05.020} {\bibfield  {journal} {\bibinfo
  {journal} {Phys. Lett.}\ }\textbf {\bibinfo {volume} {B723}},\ \bibinfo
  {pages} {257} (\bibinfo {year} {2013}{\natexlab{b}})},\ \Eprint
  {http://arxiv.org/abs/1303.1596} {arXiv:1303.1596 [astro-ph.CO]} \BibitemShut
  {NoStop}%
\bibitem [{\citenamefont {Li}\ \emph {et~al.}(2015)\citenamefont {Li},
  \citenamefont {Lin}, \citenamefont {Wang},\ and\ \citenamefont
  {Chang}}]{Li:2015uda}%
  \BibitemOpen
  \bibfield  {author} {\bibinfo {author} {\bibfnamefont {X.}~\bibnamefont
  {Li}}, \bibinfo {author} {\bibfnamefont {H.-N.}\ \bibnamefont {Lin}},
  \bibinfo {author} {\bibfnamefont {S.}~\bibnamefont {Wang}}, \ and\ \bibinfo
  {author} {\bibfnamefont {Z.}~\bibnamefont {Chang}},\ }\href {\doibase
  10.1140/epjc/s10052-015-3380-2} {\bibfield  {journal} {\bibinfo  {journal}
  {Eur. Phys. J.}\ }\textbf {\bibinfo {volume} {C75}},\ \bibinfo {pages} {181}
  (\bibinfo {year} {2015})},\ \Eprint {http://arxiv.org/abs/1501.06738}
  {arXiv:1501.06738 [gr-qc]} \BibitemShut {NoStop}%
\bibitem [{\citenamefont {{Schombert}}\ and\ \citenamefont
  {{McGaugh}}(2014)}]{2014PASA...31...11S}%
  \BibitemOpen
  \bibfield  {author} {\bibinfo {author} {\bibfnamefont {J.~M.}\ \bibnamefont
  {{Schombert}}}\ and\ \bibinfo {author} {\bibfnamefont {S.}~\bibnamefont
  {{McGaugh}}},\ }\href {\doibase 10.1017/pasa.2014.2} {\bibfield  {journal}
  {\bibinfo  {journal} {PASA}\ }\textbf {\bibinfo {volume} {31}},\ \bibinfo
  {eid} {e011} (\bibinfo {year} {2014})},\ \Eprint
  {http://arxiv.org/abs/1401.0238} {arXiv:1401.0238} \BibitemShut {NoStop}%
\bibitem [{\citenamefont {Li}\ \emph {et~al.}(2018)\citenamefont {Li},
  \citenamefont {Lelli}, \citenamefont {McGaugh},\ and\ \citenamefont
  {Schormbert}}]{Li:2018tdo}%
  \BibitemOpen
  \bibfield  {author} {\bibinfo {author} {\bibfnamefont {P.}~\bibnamefont
  {Li}}, \bibinfo {author} {\bibfnamefont {F.}~\bibnamefont {Lelli}}, \bibinfo
  {author} {\bibfnamefont {S.}~\bibnamefont {McGaugh}}, \ and\ \bibinfo
  {author} {\bibfnamefont {J.}~\bibnamefont {Schormbert}},\ }\href@noop {} {\
  (\bibinfo {year} {2018})},\ \Eprint {http://arxiv.org/abs/1803.00022}
  {arXiv:1803.00022 [astro-ph.GA]} \BibitemShut {NoStop}%
\end{thebibliography}
\end{document}